\documentstyle[preprint,aps]{revtex}
\newcommand{\lsim}{\mathrel{\mathop{\kern 0pt \rlap
  {\raise.2ex\hbox{$<$}}}
  \lower.9ex\hbox{\kern-.190em $\sim$}}}
\newcommand{\gsim}{\mathrel{\mathop{\kern 0pt \rlap
  {\raise.2ex\hbox{$>$}}}
  \lower.9ex\hbox{\kern-.190em $\sim$}}}

\newcommand{\beq}     {\begin{equation}}
\newcommand{\eeq}     {\end{equation}}

\begin{document}
\draft
\preprint{\vbox{\hbox{AS-ITP-2001-011}
}}

\title{Confronting Heavy Tau Neutrinos with Neutrino Oscillations}

\author{Chun Liu}

\vspace{1.5cm}

\address{
Institute of Theoretical Physics, Chinese Academy of Sciences,\\
P.O. Box 2735, Beijing 100080, China
}

\maketitle
\thispagestyle{empty}
\setcounter{page}{1}
\begin{abstract}
   If the tau neutrino is as heavy as $10$ MeV which may have certain
astrophysical implications, the neutrino mass pattern is studied so as to 
accommodate the new oscillation observations.  It predicts that the electron
neutrino has Marjorana mass around $0.05$ eV.  A supersymmetric model is 
described to realize the above scenario.  
\end{abstract}
\vspace{1.5cm}

\hspace{1.1cm}Keywords: tau neutrino, neutrino oscillation.


\newpage

There are some motivations for a heavy $\nu_\tau$.  Cosmologically, $10$ MeV 
$\nu_\tau$'s can compose the cold dark matter in a scenario with low re-heating
temperature \cite{giudice}.  Theoretically, a $10$ MeV $\nu_\tau$ is predicted
in a supersymmetric (SUSY) model which understands the muon mass from the 
sneutrino vacuum expectation values \cite{dls}.  One astrophysical implication 
is that gamma ray bursts may be just the supernova explosions \cite{ls}.  In 
this model, $\nu_\tau$ mixes with other neutralinos slightly.  It decays to 
light gravitino and photon with a very long lifetime $\sim 10^{13}$ sec. 
Therefore, distant supernova explosions which emit tau neutrinos look like 
gamma ray bursts to us.  

Neutrino oscillation observations should be considered carefully.  The
Super-Kamiokande (Super-K) data for the atmospheric neutrino problem (ANP) 
imply that the $\nu_\mu$ maximally mixes with $\nu_x$ ($x\neq e$) with 
$\Delta m_{\mu x}^2\simeq 3 \times 10^{-3}$ eV$^2$ \cite{superk}. And it is
claimed that compared to the sterile neutrino, the $x=\tau$ case is favored 
\cite{superk1}.  However, Ref. \cite{foot} has argued that this claim is not
yet reliable, and more careful analysis is needed.  Nevertheless, as
emphasized in Ref. \cite{gmp}, the $x=sterile$ case is not ruled out on its
own basis.  

The Sudbury Neutrino Observatory (SNO)'s first result \cite{sno} for the
solar neutrino problem (SNP) makes it clear that mixing among the active
neutrinos are essential, although certain involvement of a sterile neutrino
cannot be excluded \cite{bmw}.  Recent Super-K's result \cite{superk2} for the
SNP shows that the solutions lie in the large mixing angle (LMA) region with
$\Delta m^2 \simeq 10^{-5}-10^{-4}$ eV$^2$ or 
$\Delta m^2 \simeq 10^{-9}-10^{-7}$ eV$^2$.  

In this Letter, we take $\nu_\tau$ to be heavy ($\simeq 10$ MeV).  The ANP is
explained by introducing a sterile neutrino $\nu_s$.  The SNP is thus mainly
due to the $\nu_e-\nu_\mu$ mixing.  Can this scenario be consistent with
neutrino oscillations in detail?  There are three light neutrinos, $\nu_e$,
$\nu_\mu$ and $\nu_s$.  It looks similar to the case of three light active
neutrinos which have several forms of the neutrino mass matrix allowed by the
neutrino oscillations \cite{af}.  However, careful consideration shows that
the neutrino mass matrix is almost unique.  By introducing a sterile neutrino,
naively the pseudo-Dirac mechanism would be expected for the ANP.  But this
can not explain the SNP, because even if $\nu_e$ is taken to be degenerate
with $\nu_\mu$ and $\nu_s$, the large mixing between $\nu_e$ and $\nu_\mu$ can
not be achieved.  Requiring parameter tunings to be small, we come up with the
following neutrino mass matrix phenomenologically. It in the ($\nu_e$,
$\nu_\mu$, $\nu_s$) basis to the leading order is 
\beq
\label{1} 
{\mathcal M}_\nu^{(0)} = \frac{m}{\sqrt{2}} \left(
\begin{array}{ccc}
0 & 1 & 1 \\
1 & 0 & 0 \\
1 & 0 & 0
\end{array}
\right)\,.
\eeq
Note that the matrix elements ($12$) and ($13$) are not necessarily equal. 
The equality will be exact if the ANP is due to a maximal mixing.  

The following mass spectrum is obtained from Eq. (\ref{1}),  
\beq 
\label{2}
m_1=m_2=m, ~~~~~m_3=0\,,
\eeq
Two neutrinos are degenerate and one massless.  Their mixing matrix is then 
\beq
\label{3}
U = \left(
\begin{array}{ccc}
\frac{1}{\sqrt{2}} & -\frac{1}{\sqrt{2}} & 0                  \\
\frac{1}{2}        & \frac{1}{2}         & \frac{1}{\sqrt{2}} \\
\frac{1}{2}        & \frac{1}{2}         & -\frac{1}{\sqrt{2}}
\end{array}
\right)\,.
\eeq
The charged lepton mass matrix has been taken to be diagonal at the leading
order.  Therefore both $\nu_\mu-\nu_s$ and $\nu_e-\nu_\mu$ mixing have been
fixed to be maximal already, for the ANP and SNP, respectively.  

The value of $m$ is determined by the ANP, 
\beq
\label{4}
m\simeq 0.05 ~{\rm eV}\,.
\eeq
The degeneracy of $\nu_1$ and $\nu_2$ has to be lift as required by the SNP. 
Phenomenologically, a small perturbation $m\epsilon$ can be added to the mass
matrix Eq. (\ref{1}).  It splits $\nu_1$ and $\nu_2$ with
\beq
\label{5}
\Delta m^2_{12} \simeq m^2\epsilon \,.
\eeq
$\epsilon\simeq 10^{-2}$ is for the Mikheyev-Smirnov-Wolfenstein \cite{msw}
solution and $\epsilon\simeq 10^{-6}$ for the LOW solution.  

The fact that the SNP is due to a large mixing instead of a maximal mixing is
explained by considering the mixing matrix of charged leptons which was taken
as unit matrix at the leading order.  It is then natural to expect the
$\nu_e-\nu_\mu$ mixing angle diviates from $\pi/4$, 
$\sin 2\theta_{e\mu}\simeq 1-\sqrt{\frac{m_e}{m_\mu}}\simeq 0.93$.  

Let us discuss a SUSY model \cite{dls,c} which can produce the neutrino mass 
matrix Eq. (\ref{1}).  The model is a SUSY extension of the standard model.  
Lepton number violation is introduced so that one of the left-handed sneutrino 
gets a non-vanishing vacuum expectation value, $\langle v_3\rangle\sim$ few GeV 
which results in a $10$ MeV Majorana mass for the tau-neutrino \cite{dls}.  In 
addition, we introduce two heavy ($N_1$ and $N_2$) and one massless ($N_3$) 
right-handed neutrino superfields.  The relevant superpotential for $N_1$ and 
$N_2$ is generally written as 
\beq
\label{6}
{\cal W}_{(1,2)}\sim L_1H_uN_{1,2}+L_2H_uN_{1,2}+L_3H_uN_{1,2}
+M_1N_1N_1+M_2N_2N_2, 
\eeq
where $L_i$ ($i=1,2,3$) are the SU(2) doublet superfields of leptons, $H_u$ 
denotes one of the Higgs fields, and $M_{1,2}$ are the masses of $N_{1,2}$.  
The basis where the mass matrix of charged leptons is diagonal, is expanded by  
\beq
\begin{array}{ccc}
\label{7}
L_e    &=&\frac{1}{\sqrt{6}}(L_1+L_2-2L_3),\\
L_\mu  &=&\frac{1}{\sqrt{2}}(L_1-L_2),\\
L_\tau &=&\frac{1}{\sqrt{3}}(L_1+L_2+L_3).
\end{array}
\eeq 
In this basis, 
\beq
\label{8}
{\cal W}_{(1,2)}\sim L_eH_uN_{1,2}+L_\mu H_uN_{1,2}+L_\tau H_uN_{1,2}
+M_1N_1N_1+M_2N_2N_2.
\eeq
One observation is that the term containing $L_\tau$ is useless, because 
$\nu_\tau$ is already much heavier.  It is natural to expect that the mass 
sub-matrix of $\nu_e$ and $\nu_\mu$ in Eq. (\ref{1}) comes from the seesaw 
mechanism given by the other terms in the superpotential Eq. (\ref{8}), and 
$m\sim \langle H_u\rangle ^2/M$.

We assume $N_3$ couples to $L_3$ dominantly,
\beq
\label{9}
{\cal W}_3=cL_3H_uN_3 
\eeq 
with coupling $c$ being very small, $c\langle H_u\rangle\sim 10^{-1}-10^{-2}$ 
eV.  The smallness of $c$ can be understood if $N_3$ is a composite particle 
\cite{composite}.  In the basis of $L_e$, $L_\mu$ and $L_\tau$, 
\beq
\label{10}
{\cal W}_3\sim L_eH_uN_3 + L_\tau H_uN_3 .
\eeq
Again, the second term is not important for neutrino masses.  The first term 
just generates the $(13)$ entry of the mass matrix Eq. (\ref{1}).  Note that 
$N_3$ is almost massless and there is no $L_\mu-N_3$ coupling.  Therefore the 
texture of neutrino mass matrix Eq. (\ref{1}) is obtained.  

One phenomenologically interesting point of introducing light $N_3$ in this 
model is that it interacts with other leptons as 
\beq
\label{11}
{\cal W}'\sim \frac{1}{M}L_eL_\mu E^c_\mu N_3, 
\eeq 
where $E^c$ denotes the SU(2) singlet charged lepton superfield.  Because 
$\langle v_3 \rangle\neq 0$, the above superpotential results in the following 
interaction, 
\beq
\label{12}
{\cal L}'=\frac{v_3}{M} \mu^+ \mu^- \phi_{N_3}, 
\eeq 
where $\phi_{N_3}$ is the scalar component of $N_3$.  Note that after SUSY 
breaking, $\phi_{N_3}$ becomes massive.  So it decays to $\mu^+\mu^-$ with 
possibly a long decay lenth.  We wonder if this is related to the recent 
observation of NuTeV Collaboration \cite{nutev}, or can be tested in future 
experiments.

Experimentally, the neutrino mass scenario in this paper can be tested in the 
future.  Besides the direct measurement of $\nu_\tau$ mass, the confirmation of 
the $\nu_\mu-\nu_\tau$ oscillation for the ANP will be a serious challenge.  It 
is predicted that the electron neutrino has a Marjorana mass around $0.05$ eV.  
The neutrino-less double $\beta$-decay experiments will probe this value 
\cite{dbd}.  There is no room for LSND result, but it is compatible with KARMEN 
experiment.  The mixing $U_{e3}$ can be vanishingly small without affecting the 
physics discussed in this paper.

\vspace{1.5cm}
 
When the work was written, we got to know Ref. \cite{habig} which reports
$\tau$ appearence in the ANP observation at 2$\sigma$ level.  The author is
supported in part by the National Natural Science Foundation of China with
grant no. 10047005.

\end{document}